
\input harvmac.tex

\input epsf
\epsfverbosetrue
\def\fig#1#2{\topinsert\epsffile{#1}\noindent{#2}\endinsert}
\def\fig#1#2{}
%
\def\Title#1#2{\rightline{#1}
\ifx\answ\bigans\nopagenumbers\pageno0\vskip1in%
\baselineskip 15pt plus 1pt minus 1pt
\else
\def\listrefs{\footatend\vskip 1in\immediate\closeout\rfile\writestoppt
\baselineskip=14pt\centerline{{\bf References}}\bigskip{\frenchspacing%
\parindent=20pt\escapechar=` \input
refs.tmp\vfill\eject}\nonfrenchspacing}
\pageno1\vskip.8in\fi \centerline{\titlefont #2}\vskip .5in}

\ifx\answ\bigans\def\tcbreak#1{}\else\def\tcbreak#1{\cr&{#1}}\fi
\message{If you do not have msbm (blackboard bold) fonts,}
\message{change the option at the top of the tex file.}
\font\blackboard=msbm10 scaled \magstep1
\font\blackboards=msbm7
\font\blackboardss=msbm5
\newfam\black
\textfont\black=\blackboard
\scriptfont\black=\blackboards
\scriptscriptfont\black=\blackboardss

%
\def\yboxit#1#2{\vbox{\hrule height #1 \hbox{\vrule width #1
\vbox{#2}\vrule width #1 }\hrule height #1 }}
\def\fillbox#1{\hbox to #1{\vbox to #1{\vfil}\hfil}}
\def\ybox{{\lower 1.3pt \yboxit{0.4pt}{\fillbox{8pt}}\hskip-0.2pt}}
\def\comments#1{}

\def\p{\partial}

\def\tr{{\rm tr\ }}

\Title{\vbox{\baselineskip12pt
\hfill{\vbox{
\hbox{BROWN-HET-993 \hfil}
\hbox{UT-Komaba 95/5 \hfil}
\hbox{hep-th/9504091}}}}}
{\vbox{\centerline{Doubling of Scattering Phase Shifts}
\vskip20pt
\centerline{for Two Dimensional Strings}}}
\centerline{Antal Jevicki\quad\qquad Miao Li}
\smallskip
\centerline{Department of Physics}
\centerline{Brown University}
\centerline{Providence, RI 02912}
\centerline{\tt antal@het.brown.edu \quad li@het.brown.edu}
\vskip10pt
\centerline{Tamiaki Yoneya}
\centerline{Institute of Physics}
\centerline{University of Tokyo}
\centerline{Komaba, Tokyo 153}
\centerline{\tt yoneya@tansei.cc.u-tokyo.ac.jp}
\bigskip
We discuss the origin of the leg factors appearing in 2D string theory.
Computing in the world sheet framework we use the semiclassical method to
study string amplitudes at high energy. We show that in the case of a simplest
2-point amplitude these factors correspond entirely
to the time delay for reflection off the Liouville wall. Our
semiclassical calculation
reveals that the string longitudinal modes, although nonpropagating in
2D spacetime, have the effect of doubling the phase shift. Particular
emphasis is put on comparison with the point particle (center of mass) case.
A general method is then given for calculating an arbitrary amplitude
semiclassically.
\noindent

\Date{April 1995}
\nref\review{I. Klebanov, String Theory in Two Dimensions, in
``String Theory and Quantum Gravity'', Proceedings of the Trieste
Spring School 1991, eds. J. Harvey et al. (World Scientific, Singapore, 1991);
A.~Jevicki, Developments in 2D String Theory, BROWN-HET-918, hep-th/9309115;
P.~Ginsparg and G.~Moore, Lectures on 2D Gravity and 2D String Theory,
YCTP-P23, LA-UR-92-3479; J.~Polchinski, What is String Theory, NSF-ITP-94-97,
hep-th/9411028.}
\nref\polyakov{A. Polyakov, Mod. Phys. Lett. A6 (1991) 635;
D.~Gross, I.~Klebanov and M.~Newman, Nucl. Phys. B359 (1991) 621;
D.~Minic and Z.~Yang, Phys. Lett. B274 (1992) 27, hep-th/9109012; N.~Sakai and
Y.~Tanii, Phys. Lett. B276 (1992) 41, hep-th/9108027.}
\nref\np{M. Natsuume and J. Polchinski, Nucl. Phys. B424 (1994) 137,
hep-th/9402156; J.~Polchinski, Phys. Rev. Lett. 74 (1995) 638, hep-th/9409168.}
\nref\jy{A. Jevicki and T. Yoneya, Nucl. Phys. B411 (1994) 64, hep-th/9305109.}
\nref\gl{M. Goulian and M. Li, Phys. Rev. Lett. 66 (1991) 2051.}
\nref\mss{G. Moore, N. Seiberg and M. Staudacher, Nucl. Phys. B362
(1991) 665.}
\nref\dvv{R. Dijkgraaf, H. Verlinde and E. Verlinde, Nucl. Phys. B371 (1992)
269.}
\nref\joe{J. Polchinski, Nucl. Phys. B346 (1990) 253.}
\nref\soliton{R. Dashen, B. Hasslacher and A. Neveu, Phys. Rev. D11 (1975)
271; J.-L.~Gervais
and A.~Jevicki, Nucl. Phys. B110 (1976)  93; 113; L.~D.~Faddeev and
V.~E.~Korepin, Phys. Rep. 42 (1978) 1, and references therein.}
\nref\gn{J.-L. Gervais and A. Neveu, Nucl. Phys. B238 (1984) 125.}
\nref\ct{T. L.  Curtright and C. B. Thorn, Phys. Rev. Lett. 48 (1982) 1309.}
\nref\bct{E. Braaten, T. Curtright and C. Thorn, Ann. Phys. 147 (1983) 365.}
\nref\seiberg{N. Seiberg, Notes on Quantum Liouville Theory and Quantum
Gravity, Prog. Theor. Phys. Suppl. 102 (1990) 319.}
\nref\bg{A. Bilal and J.-L. Gervais, Nucl. Phys. B305 (1988) 33.}
\nref\ven{D. Amati, M. Ciafoloni and G. Veneziano, Phys. Lett. B197 (1987) 81.}
\nref\gm{D.  Gross and P. Mende, Phys. Lett. B197 (1987) 129; Nucl. Phys.
B303 (1988) 407.}
\nref\yo{T. Yoneya, "{\it Wandering in the Fields}",
(eds. K. Kawarabayashi and A. Ukawa, World Scientific, Singapore, 1987) p. 419;
Mod. Phys. Lett. A4 (1989) 1589. }
\newsec{Introduction}

It is generally believed that the $c=1$
matrix model describes a critical string theory in a two-dimensional
target spacetime with coordinate dependent condensation of dilaton
and tachyon fields \review. The scaling properties of the
matrix model observables with respect to the fermi energy agree
precisely with the
predictions of the string theory, when suitable correspondences
are made with the physical variables of string theory.
In particular, the string $S$-matrix coincides with the
$S$-matrix of the collective-field quanta of the
matrix model after making a momentum dependent
renormalization of the asymptotic fields. The renormalization
factor, often called the leg factor, is a
pure phase, but cannot simply
be discarded as unobservable since it contributes to
a momentum dependent phase shift of the scattering amplitudes.
The leg factors contain important physical
information on the nature and content of two dimensional string theory. When
continued to Euclidean
metric, they exhibit an infinite number of poles
whose positions coincide with the characteristic
values of momenta of 2D strings. Namely, the poles
occur at the momenta where energy-momentum conservation law
is obeyed in the linear-dilaton background
up to, in general, a finite number of insertions of the background
tachyon fields \polyakov.
If one first neglects the tachyon condensation,
one can easily perform an ordinary free field calculation of
the $S$-matrix for bulk scattering.
Then, the poles can be interpreted as
arising from the short distance singularities of the product
of the vertex operators on the world sheet. The short-distance
sigularities on the world sheet correspond to large distance propagation
of strings in the asymptotic region of the target spacetime.
We expect that such bulk amplitudes are, in general,
exponentially damped, because of the linear dilaton condensation.
Ref.\np\ gives a spacetime interpretation of the bulk
scattering, and shows that the first few
singularities indeed account for the exponentially small effect of
the string interactions occuring in the asymptotic region of the target
spacetime.

{}From the view point of bulk scattering, the poles reflect only the
asymptotic properties of the target spacetime and do not
reflect the presence of the tachyon condensation.
However, the same leg factor simultaneously
accounts for the amplitudes in scattering (so-called {\sl wall}
scattering)
against the tachyon background. In particular, the two-point
amplitude extracts the response of the string
propagation upon the tachyon background. Here,
the poles should be interpreted as arising solely from the resonance
of strings with a coordinate dependent tachyon background.

It is not at all clear, however,
why the same leg factor can explain both of these
properties of 2D strings which seem to be independent to each
other. Is it just a coincidence
valid only for a particular background? Or, is there anything
deeper, related to unknown universal properties of string theories?
This is, we believe, one of crucial questions to be clarified in seeking
a nonperturbative and background independent
formulation of string theories, based on
a hint provided by the matrix models. The problem becomes
serious especially when we consider possible matrix models
corresponding to different string backgrounds, since
the pole structure implied from the resonance with the
background does not completely coincide the
pole structure required for bulk scattering.
For a discussion on this problem and a tentative proposal in the case
of large-mass black hole background, see \jy.

In view of this situation, the purpose of the present paper is
to give a new direct derivation of
the scattering phase shifts at high energy
from the view point of continuum string theory.
Although our results are still too modest to answer the above questions,
we hope that they enlighten some aspects
that have not been fully recognized in the previous works.

To our knowledge, no string theoretic derivation of the
wall scattering amplitude has been given
except for the one based on the method
of analytical continuation with respect to
the number of insertions of the tachyon condensate operators \gl.
Although such a derivation nicely gives a
general n-point amplitude in closed form, it does not reveal the
physical origin of the leg factors.

We shall consider the high energy limit of the wall scattering
in order to probe inside the wall region of the target spacetime.
Throughout the present paper, we remain in the lowest approximation
with respect to the string coupling constant by assuming
a strong tachyon condensation.
We use a direct semi-classical approximation and explicitly
compute the effect of higher string modes
in the tachyon background.
We then find that string fluctuations around the
center of mass give an equal contribution to the phase shift
as that coming from the center of mass
motion itself. This result is consistent with the double pole
structure of the two-point amplitude and suggests that the leg factor
not only reflects the properties of the asymptotic
propagation, but also embodies a property originated
from the intrinsically extended nature of 2D string in a given
background. Although there are no transverse oscillations, the
extension in the longitudinal direction does give an observable
effect. This should perhaps be interpreted as a manifestation of
discrete physical states of 2D string theories.

In the next section, we begin by briefly reviewing
2D string theory with tachyon and linear-dilaton background.
It will be emphasized that a simple reduction to the
center of mass motion is not sufficient to account for the
phase shift of the 2-point amplitude, especially in the
high energy limit. In section 3, we will proceed to calculate
the effect of higher string modes and compute the contribution
of them to the phase shift in the high-energy limit.
We first present a calculation based on a direct mode counting
in Minkowski metric.  Next in section 4,
we propose a general scheme for calculating
arbitrary amplitudes in the semi-classical approximation
in Euclidean space. The result for the two-point
amplitude obtained in section 3 is reproduced by this scheme.
In the concluding section, we will briefly discuss the spacetime meaning
of the phase shift and try to give a qualitative interpretation of
our result.

\newsec{Leg Factors of the Scattering Amplitudes}

In this section we present a brief review of 2d string theory, introduce
the notation and describe some earlier attempts at understanding the origin
and pole structure of its S-matrix.

In the world sheet description of the theory one has $c_M=1$ matter
$X (z,\bar{z})$ together with a $c_L = 25$ Liouville field $\varphi
(z,\bar{z})$:
\eqn\worsh{
S = {1\over 8\pi} \int d^2z \sqrt{{\hat g}} \left\{ \hat{g}^{ab}
\partial_a X \partial_b X + \hat{g}^{ab} \partial_a \varphi \partial_b
\varphi + Q \, \hat{R}^{(2)} \varphi + \mu \, e^{\alpha \varphi }
\right\},}
with $Q = - 2 \sqrt{2} , \alpha = \alpha_- = - \sqrt{2}$.  The vertex
operators  for massless tachyons read
\eqn\vertex{
V ( z , \bar{z} ) = e^{i\sqrt{2}  kX(z,\bar{z} )} \, e^{- \sqrt{2} (1-
\vert k \vert ) \varphi (z , \bar{z} )},}
where one only takes the (physical) + dressing of the Liouville field.
In the above, we have the Euclidean theory, a transformation:
\eqn\transf{
iX \rightarrow X \quad , \quad k \rightarrow \pm i\omega,}
(with $\omega > 0$ ) brings us to the Minkowski time and the vertex operators
\eqn\minkow{
T_{\pm \omega} = e^{- i\sqrt{2} \omega X (z , \bar{z} )} \, e^{-\sqrt{2}
(1 \pm i\omega ) \varphi (z , \bar{z} )}}
describing respectively left and right moving waves.

The $n$-point tachyon $S$-matrix was found through a combination of
continuum and matrix model techniques to have the general structure \review
\eqn\ampli{
S (k_1 , k_2 \cdots k_n ) = \prod_i {\Gamma (-2\vert k_i \vert )\over
\Gamma ( 2\vert k_i \vert )} \, \bar{S} (k_1 , k_2 , \cdots k_n ).}

It is characterized by a leg factor for each scattering particle and a
residual  $S$-matrix  $\bar{S}$ described by the simple dynamics of a
collective field.  The origin and physical interpretation of the leg factors
however is more mysterious.  They contain poles at imaginary values of
momenta, these poles partially come from intermediate discrete states which
are known to appear in the  spectrum of 2d String Theory.  But the poles
also come from the nontrivial tachyon background described by the Liouville
exponential potential $\int \mu \, e^{-\sqrt{2} \varphi (z , \bar{z} )}$.
The latter is especially clear if one takes the  simplest example of the
2-point $S$-matrix.  In this case one has a single string reflecting of the
Liouville wall with the amplitude
\eqn\twop{
S_2 (\omega , -\omega ) = \mu^{-2i\omega} \left( {\Gamma (2i\omega )\over
\Gamma (-2i\omega )} \right)^2.}
This amplitude exhibits \underbar{double} poles at $2i\omega_n = -n$.  Since
this is a one string process, these poles can not come from typical exchange
states associated with intermediate Feynman propagators.  They entirely
reflect the property of the background and the fact that the object being
scattered is a string.

In this connection one can consider a point particle which would correspond
to  the center of mass (or the zero mode) of the string.  The
scattering of the Liouville wall is given by the Klein-Gordon (Wheeler-de Witt)
equation \mss\dvv:
\eqn\wave{
\left( {\partial^2\over \partial t^2 } - {\partial^2\over \partial
\varphi_{0}^{2}} + \mu \, e^{-\sqrt{2} \varphi_{0} } \right) \psi (t,
\varphi_0 ) = 0,}
which is solved to give the scattering amplitude
\eqn\single{
T(\omega ) = {\Gamma(2i\omega )\over \Gamma (-2i\omega )} \, \mu^{-2i\omega}
= e^{i\delta (\omega )}.}
This amplitude  is again a pure phase but exhibits single imaginary energy
poles of $-2i\omega = n$.  These clearly come from the Liouville wall
$e^{-\sqrt{2} \varphi_{0}}$ itself.  It has been suggested that the center of
mass Wheeler-de Witt equation might be capable of giving an exact description
of string theory.  For that scenario to be true, one has to come up with a
mechanism for generating the double poles of the string $S$-matrix.
Ref. \np\  has made a suggestion of using a modified tachyon background of
the form
\eqn\logp{
T_0 (\varphi_0 ) = (b_1 + b_2 \varphi ) \, e^{-\sqrt{2} \varphi}.}
The modification given by the second term is capable of converting a single
into a double pole
\eqn\double{
\langle k \vert \varphi_0 \, e^{-\sqrt{2} \varphi_{0}} \vert k \rangle =
{1\over (2ik+1)^2}}
at least for the lowest case.
The modified background was argued by Polchinski \joe\  as arising in the
effective tachyon Lagrangian
\eqn\effective{
L_{eH} = {1\over 2g_{st}^2} \int d^2 x \sqrt{-G} \, e^{-2D} \left[ -
\left( \nabla T\right)^2 + 4T^2 - V(T) + a \left\{ R + 4 \left( \nabla D
\right)^2 + 16 \right\} + \cdots \right],}
where it corresponds to a  static solution for the tachyon field $T_0
(\varphi_0 ) $.  The value of the coefficients $b_1$ and $b_2$ can in
principle be specified by the nonlinear interaction term of the effective
Lagrangian or the full string field theory.  In the continuum, it has been
suggested that a similarly modified Liouville term could be generated through
the fact that for $c<1$ one actually has two possible candidates
$$e^{\alpha_{\pm} \varphi} \qquad {\rm with} \qquad \alpha_{\mp} = -
\sqrt{2} \mp \epsilon $$
as $c\rightarrow 1$.  Forming a linear combination and taking the limit
$\epsilon\rightarrow 0$ we can then  have :
\eqn\limit{
{1\over 2\epsilon} \left[ e^{\alpha + \varphi} - e^{\alpha - \varphi}\right]
\rightarrow
\varphi (z , \bar{z} ) \, e^{-\sqrt{2} \varphi (z , \bar{z} )}.}

The problem with the above mechanism for generating double poles is that it
appears to be at best relevant only for low momenta.  It does not survive at
high energy where it would require the addition of an infinite number of
further terms.  It is then more appropriate to concentrate on the high
energy region in trying to discover the actual mechanism for doubling
of poles in the string $S$-matrix.
At high energy this  means a doubling of the scattering phase shift and the
time delay.  As we have mentioned, a simple
modification by an additional term of the form $\varphi e^{-\sqrt{2}\varphi}$
can not achieve this at high momenta as is easily seen
in the classical-particle picture: Near the turning point
$e^{-\sqrt{2}\varphi + \log \varphi} \sim e^{-\sqrt{2}\varphi}$
as $\varphi\sim -\ln \omega \rightarrow -\infty$ and the prefactor $\varphi$
does not play any leading role in the high-energy limit.
{}.

Typical high energy scattering can be studied by semiclassical techniques
and that is what we do in the next section.  In the mechanism that we
suggest, a central role is played by the higher string modes.  In an expansion

\eqn\modes{
X_{\mu} (\sigma) = X_{\mu}^0 + \sum_{n\not= 0} \, e^{in\sigma } X_{\mu}^{(n)},}
the non-zero string modes will be shown to double the contribution to the
phase shift. Such effects of higher modes participating to renormalize the
leading classical result were known to systematically appear in
integrable two-dimensional soliton models \soliton. Their presence
in Liouville theory has been emphasized originally in \gn\ and \ct. Our
discussion concerns the evaluation of a physical quantity ( the string
S-matrix) at high energies with a demonstration of a significant one loop
effect doubling the classical contribution.

\newsec{ Semi-classical Calculation in Minkowski Space}

It is argued in the previous section that the degree of the center of mass
of a string alone does not account for the phase shift of the 2-point
scattering amplitude. We now set out to show this is indeed true
by an explicit computation. For the 2-point amplitude, it is more
physical to work with a Minkowski spacetime, and for this matter to
work with a Minkowski world sheet also.  The amplitude is then
expressed as a world sheet path integral with appropriate boundary
conditions set by wave functions at $t=\pm\infty$, the world sheet
topology is that of a cylinder, as we consider closed string only.
There is a systematic quantization procedure developed in \bct.
Unfortunately
we have found it difficult to adapt that procedure to the calculation
of string scattering amplitudes. Therefore,
we shall perform only a semi-classical calculation, in which the zero mode
of $\varphi$ is taken as the one satisfying the classical Liouville
equation, and the one-loop calculation is done by expanding around this
classical solution.
It turns out that the semi-classical computation on the world sheet
becomes very accurate at high energies.

The Minkowski world sheet action with a  Minkowski target space is
\eqn\mink{S(\varphi,X)={1\over 8\pi}\int dtd\sigma\left((\p\varphi)^2
-(\p X)^2-\mu e^{-\sqrt{2}\varphi}\right),}
where $X$ is the real time of the target Minkowski space and $t$ is the World
sheet time. The period of world sheet space $\sigma$ is $2\pi$. Consider
the 2-point amplitude. The incoming wave is left-moving, so at $t=-\infty$,
we shall insert wave function $T_\omega=\exp\left(-i\sqrt{2}\omega(X+\varphi)
\right)$, this determines the boundary condition of the path integral at
$t=-\infty$. The out-going wave is right-moving, therefore an insertion
of the complex conjugate of $T_{-\omega}=\exp\left(-i\sqrt{2}\omega(X-\varphi)
\right)$ is used. Note that in the wave functions a factor $\exp
(-\sqrt{2}\varphi)$ is dropped out, since this factor is cancelled by
appropriate boundary term in the action, whose origin is the background
charge term. Now the 2-point amplitude is given by
\eqn\twopoint{S_2(\omega,-\omega)=\int [dXd\varphi]\overline{T}_{-\omega}
(X(\infty),\varphi(\infty))T_\omega(X(-\infty),\varphi(-\infty)) e^{iS}.}
The integral over $X$ is Gaussian, thus can be easily performed. The
insertions of wave functions only constrain $\p_tX(\pm\infty)\sim
\omega$. Their effect is then removed by a shift $X\rightarrow
2\sqrt{2}\omega t +X$.

The nontrivial part of the computation concerns integration over $\varphi$.
Let us separate $\varphi$ into a (center of mass) zero mode and oscillating
modes $\varphi=\varphi_0(t)+\varphi_{os}$. The total action for $\varphi$,
taking the boundary wave functions into account, is
\eqn\tot{S(\varphi)={1\over 8\pi}\int dtd\sigma\left((\p_t\varphi)^2
-(\p_\sigma\varphi)^2-\mu e^{-\sqrt{2}\varphi}\right)-\sqrt{2}\omega
(\varphi(\infty)+\varphi(-\infty)).}
It follows from variation of the above action that the Liouville zero mode
satisfies the Liouville equation
\eqn\liou{\p_t^2\varphi_0-{\mu\over\sqrt{2}}e^{-\sqrt{2}\varphi_0}=0}
with the boundary conditions
$$\p_t\varphi_0(\pm\infty)=\pm 2\sqrt{2}\omega.$$
The solution is well-known \seiberg:
\eqn\solut{e^{-\sqrt{2}\varphi_0}={8\omega^2\over\mu \cosh^2 2\omega t}.}
Plugging this solution into \tot, we will get an infinite contribution.
The divergence is proportional to the time lapse and therefore should be
subtracted in order to obtain the true phase shift. Perhaps the simplest way
to calculate the classical action is to vary the action with respect
to $\omega$, then integrate the result to get the zero mode (or center
of mass) contribution to the phase shift. Vary the action with respect
to $\omega$ and make use of the Liouville equation and the boundary
conditions
$$\eqalign{\delta S_{cl}&=-\sqrt{2}(\varphi_0(\infty) +\varphi_0
(-\infty)) \delta\omega\cr
&=2\ln ({8\omega^2\over\mu})\delta\omega-4\ln(\cosh 2\omega T)\delta\omega,}
$$
where a cut-off in the time lapse $T$ was introduced. The last term in the
second equality, depending on $T$, should be dropped out. The second term,
after integrating over $\omega$, gives
\eqn\com{S_2^{cl}(\omega,-\omega)=e^{iS_{cl}}=\left({\mu\over 2}
\right)^{-2i\omega}e^{i(-4\omega+4\omega\ln(2\omega))}.}
Comparing this result to that in \single, we see that they essentially
agree. The difference gets smaller and smaller when $\omega$ gets larger
and larger. This computation shows that another factor $\Gamma(2i\omega)/
\Gamma(-2i\omega)$ in the full 2-point amplitude is missing, if one takes
only the degree of center of mass into account as in \mss.

We now show that most of the other gamma ratio can be recovered from the
one-loop calculation. One-loop contribution will involve all oscillating
modes in an essential way. This suggests that even though the longitudinal
modes are nonpropagating in two dimensional spacetime, they are essential in
the full stringy description
of the scattering amplitudes (they might reflect the remnant modes, the
discrete states). Expand the action around the above classical solution
to the second order of the high modes
\eqn\expand{S(\varphi_0+\varphi)=S_{cl}+{1\over 8\pi}\int dtd\sigma
\left((\p\varphi)^2-{8\omega^2\over\cosh^2 2\omega t}\varphi^2\right).}
There are two ways to proceed to calculate the determinant resulting
from integration of high modes $\varphi$. The first method is the so-called
phase shift method. This method has the advantage of showing clearly
how high modes make a contribution $4i\omega\ln(2\omega)$. The second
method is that of heat kernel. We shall present yet a third, the
most general method in Euclidean space in the next section.

The phase shift method requires solving the eigen-value problem of the
following equation
\eqn\eigen{\left(-\p_t^2+\p_\sigma^2-{8\omega^2\over\cosh^2 2\omega t}
\right)\varphi=\lambda^2\varphi.}
The operator $\p_\sigma^2$ can be replaced by $-m^2$, its eigen-value.
It is easy to see that the equation
$$\left(-\p_t^2-{8\omega^2\over\cosh^2 2\omega t}\right)\varphi=\nu^2\varphi$$
has the solution
\eqn\doub{\varphi=e^{i\nu t}\left(\tanh 2\omega t-{i\nu\over 2\omega}\right).}
The other solution is obtained by $\nu\rightarrow -\nu$. The eigen-value
of the equation \eigen\ is then $\nu^2-m^2$, and the one-loop phase
shift is given by
\eqn\onelp{\ln Z_1=-{1\over 2}\sum_{\nu,m}\ln(\nu^2-m^2+i\epsilon).}
In order to evaluate this sum, we put equation \eigen\ into a box
$(-T,T)$. Without the background, $\nu_n$ would be $\pi n/T$. With the
background, $\nu_n=\pi n/T-\delta_n/(2T)$. $\delta_n$ is called the phase
shift (hence the name phase shift method), and is determined by
requiring $\varphi(T)=\varphi(-T)$. Using \doub\ we find
$$e^{i\delta_n}={\nu_n+2i\omega\over \nu_n-2i\omega}$$
or $\delta(\nu)=2\arctan(2\omega/\nu)$. The one-loop phase shift \onelp\
is written as
$$\eqalign{&-{T\over 2\pi}\sum_m\int d\nu(1+{\delta'(\nu)\over 2T})
\ln(\nu^2-m^2+i\epsilon)=\cr
&-{T\over 2\pi}\sum_m\int d\nu\ln(\nu^2-m^2+i\epsilon)-{1\over 4\pi}
\sum_m\int d\nu\delta'(\nu)\ln(\nu^2-m^2+i\epsilon).}$$
The first term is independent of the background, therefore of $\omega$.
It should be dropped out. The second term after integration by parts
becomes
$${1\over 2\pi}\int d\nu \delta(\nu) \sum_m {\nu\over \nu^2-m^2+i\epsilon}
=-{i\omega\over 2}\int d\nu\epsilon(\nu)\delta(\nu),$$
where $\epsilon(\nu)$ is the step function. The above equality is valid
only when $\omega>>1$. Use the previous result
for $\delta(\nu)$, the above integral over $\nu$ is logarithmically
divergent. Imposing cut-off $\Lambda$ for $\nu$ we finally
obtain the result
\eqn\correc{\ln Z_1=-4i\omega\left(1-\ln (2\omega/\Lambda)\right).}
The finite part is $-4i\omega(1-\ln(2\omega))$, just the asymptotic
value of the logarithmic of the missing ratio
$\Gamma(2i\omega)/\Gamma(-2i\omega)$.
The cut-off part $-4i\omega\ln\Lambda$ can be combined with
the classical part $-2i\omega\ln\mu$ to give a renormalized ``cosmological
constant'' $\mu\Lambda^2$. Note that this one-loop calculation is valid for
large $\omega$. An Euclidean calculation presented in the next section will
give an exact one-loop result.

The above calculation tells us the following points. First, it is
not necessary to use the operator $\varphi\exp(-\sqrt{2}\varphi)$
in the world sheet formalism as the tachyon condensate. This tachyon
condensate can be viewed at best as an effective condensate at
low energies, and perhaps
incapable of accounting for the high energy behavior. Second, high modes
play an important role even in the 2-point amplitude.
If one trusted in a naive gauge-fixing procedure, one would have
concluded that the higher modes are irrelevant.
Third, the logarithmic behavior of the amplitude is
associated with the logarithmic ultraviolet divergence of the one-loop
contribution. This shows that there is no further logarithmic
and power behaved terms from higher loops, since higher loops are
not divergent in two-dimensions. Thus, one-loop calculation gives
an exact result in the high-energy limit.

The next method employed to calculate the determinant is the heat kernel
method. This method is efficient when one tries to calculate the variation
of the determinant with $\omega$ first
\eqn\varia{\delta\ln Z_1=-{1\over 2}\delta\tr\ln\left(-\p_t^2+\p_\sigma^2
-{8\omega^2\over\cosh^22\omega t}\right)=4\int dtd\sigma
K(t\sigma,t\sigma)\delta({\omega^2\over\cosh^2 2\omega t}),}
where $K$ is the inverse of the operator $-\p_t^2+\p_\sigma^2
-8\omega^2/(cosh^2 2\omega t)$. A general method of computing $K$ was given
in the last reference in \soliton. We shall not give the derivation
except the lengthy result:
\eqn\heat{\eqalign{K(t\sigma,t\sigma)=&{i\over 4\pi}\sum_{m\ne 0} {1\over
m(a_m^2-1)}(1+{m^2\over 4\omega^2})[(1+a^2_m)(\tanh^2 2\omega t+
{m^2\over 4\omega^2})\cr
&-a_m\left((\tanh 2\omega t-{im\over 2\omega})^2+c.c\right)],\cr
a_m=&e^{2imT}{1-im/(2\omega)\over 1+im/(2\omega)},}}
again we have put the system in a box $(-T,T)$. After a straightforward
but tedious calculation, we obtain the same result as in \correc.

This second method will be most efficient in the case of semi-classical
calculation of a general amplitude, where one has to deal with Euclidean
spacetime and Euclidean world sheet. This we shall do in the next section.

\newsec{ Semi-classical Calculation in Euclidean Space}

Amplitudes are calculated on the Euclidean world sheet by inserting vertex
operators and integrating over positions of these operators. The Euclidean
world sheet action is given in eq.\worsh. In complex coordinates it reads
\eqn\worlsh{S={1\over 2\pi}\int d^2z\left(\p X\bar{\p}X+\p\varphi\bar{\p}
\varphi +\mu e^{-\sqrt{2}\varphi}\right),}
where the background charged term drops out due to the flatness of the world
sheet metric. A general amplitude is given by
\eqn\ampl{S_n(\omega_i)=\int\prod_i d^2z_i\int[dXd\varphi]e^{-S}
\prod_iV_{\omega_i}(z_i),}
with
$$V_{\omega_i}=e^{i\sqrt{2}\omega_iX-\sqrt{2}(1-|\omega_i|)\varphi}.$$
Again the $X$ part of the path integral can be easily performed, leaving
a standard correlation function together with a $\delta$ function enforcing
the conservation of energy.

Since we are considering high energy scattering, we will denote $|\omega_i|
-1$ simply by $\omega_i$, assuming that all $\omega$'s are positive. The
semi-classical computation of the $\varphi$ path integral is divided into two
steps, the same as in the previous section. The first step is to solve the
classical Liouville equation with sources provided by insertion of vertex
operators
\eqn\sour{\p\bar{\p}\varphi_0+{\mu\over\sqrt{2}}e^{-\sqrt{2}\varphi_0}
-\sqrt{2}\pi\sum_i\omega_i\delta^2(z-z_i)=0.}
This equation in principle can be solved for an arbitrary number of sources,
see \bg. For a classical calculation, it is enough to know the
behavior of the solution near each source. Near $z_i$
\eqn\asmp{\varphi_0(z)=\sqrt{2}\omega_i\ln|z-z_i|^2+{1\over\sqrt{2}}\ln\mu
+\Delta_i(\omega),}
where $\Delta_i(\omega)$ as a constant is a function of $\omega$'s, its
precise form is to be determined by the exact solution. Following \bg,
we call it the time delay at $z_i$.
\foot{This time delay should not be confused with the actual time delay
discussed in the next section.}

Again it is easy to compute the classical action including source terms by
first considering its variation with respect to $\omega$'s, as we did in
the previous section. Making use of the Liouville equation, one finds
\eqn\simp{\delta S_{cl}(\omega)=\sqrt{2}\sum_i\varphi_0(z_i)\delta\omega_i.}
Once again this is divergent since $\varphi_0(z_i)$ is divergent. The
divergence, similar to the one in the Minkowski calculation which
depends on the
time lapse, is regularized by introducing a short-distance cut-off on
the world sheet and absorbed into a renormalization of the vertex operators.
This being done, we obtain a finite result
\eqn\regu{S_{cl}(\omega)=\ln\mu\sum_i\omega_i+\sqrt{2}\int^{\omega_i}
\Delta_i(\omega)d\omega_i.}
In order for the last integral to make sense, the time delays $\Delta_i$
should satisfy the integrable conditions $\p_{\omega_i}\Delta_j
=\p_{\omega_j}\Delta_i$. This formula for the classical part of the
amplitude expresses it as integration of the time delays is due to
Bilal and Gervais \bg. We now address the question of quantum correction.

The one-loop calculation boils down to a calculation of the determinant
of operator $-\p\bar{\p}+\mu\exp(-\sqrt{2}\varphi_0)$. At the first sight,
this appears a formidable task, since for an arbitrary number of sources,
the exact solution $\varphi_0$ is rather complicated. We will see that
as far as one can calculate $\varphi_0$, the calculation of the determinant
is straightforward. Again we adopt the heat kernel method. The variation
of the logarithmic of the determinant is given by
\eqn\heatk{\delta S_1={1\over 2}\delta\ln(-\p\bar{\p}+\mu e^{-\sqrt{2}
\varphi_0})={\mu\over 2}\int d^2zK(z,z)\delta(e^{-\sqrt{2}\varphi_0}).}
The heat kernel satisfies
$$(-\p\bar{\p}+\mu e^{-\sqrt{2}\varphi_0})K(z,w)=\delta^2(z-w).$$
Suppose we know how to solve the Liouville equation \sour\ with an additional
source $\omega$ at $w$, and call this solution $\varphi_\omega(z,w)$.
It is easy to see that the following function
\eqn\deriv{-{1\over\sqrt{2}\pi}{\p\varphi_\omega(z,w)\over\p\omega}
|_{\omega=0}}
satisfies the heat kernel equation. Therefore it is to be identified
with $K(z,w)$ up to a function annihilated by the differential operator.
It follows from this observation that
$${\p\varphi_0(z_i)\over\p\omega_j}=-\sqrt{2}\pi K(z_i,z_j),$$
and since $K(z_i,z_j)$ is symmetric in $z_i$ and $z_j$, we have
$\p_{\omega_i}\Delta_j=\p_{\omega_j}\Delta_i$. Note that the above
identification of the derivative of $\varphi_0(z_i)$ with the heat kernel
is correct only up to a function which is annihilated by the Liouville
differential operator. That the integrable conditions of time delays follow
from this identification suggests that it is correct, and the function
annihilated by the Liouville differential operator is zero.

The above considerations are general. Now we demonstrate this procedure
by an explicit calculation of the 2-point function. In this case $\omega_1
=\omega_2=\omega$. Placing one source at $z=0$ and another at $z=\infty$,
the classical solution is
\eqn\clas{e^{-\sqrt{2}\varphi_0}={8\omega^2|z|^{4\omega}\over \mu(1+
|z|^{4\omega})^2}.}
Apply \regu,
\eqn\examp{S_{cl}=2\omega\ln(\mu/2)+4\omega-4\omega\ln(2\omega),}
agrees with the Minkowski calculation.

The one-loop calculation is considerably more involved. Here to obtain
the heat kernel, one has to solve the Liouville equation \sour\ with
three sources. Two sources are equal, and are the original sources placed
at $z=0$ and $z=\infty$. The third source is placed at $w$, and is taken to
zero after the derivative in \deriv\ is taken. The Liouville solution
with three sources can be expressed in terms of hyper-geometric functions,
we refer to \bg\ for details. Here it is sufficient for us to write down
the formula for $K(z,z)$, which is obtained after a lengthy calculation
\eqn\neat{K(z,z)=-2\ln|z|^2+2\left(\psi(2\omega)+2\psi(-2\omega)-2
\psi(1)\right),}
where $\psi(x)$ is the function $\Gamma'(x)/\Gamma(x)$. We remark that
the above finite result is obtained also upon certain regularization.
Plugging \neat\ into formula \heatk, we obtain the one-loop result
\eqn\quantum{S_1=\ln\left(\Gamma(-2\omega)/\Gamma(2\omega)\right)
+4\psi(1)\omega.}
The asymptotic behavior of the first term at large $\omega$ is
as desired. The last term may be interpreted as a finite renormalization
of the cosmological constant, which is regularization scheme
dependent.

We conclude that calculations in both the Minkowski world sheet
and the Euclidean world sheet correctly account for the high energy
behavior of the 2-point amplitude. While the Minkowski calculation
is more physical, and shows clearly how higher string modes modify the
zero mode contribution, the Euclidean calculation is more powerful
if one wishes to calculate high point amplitudes.

\newsec{Discussions: The Time Delay and String Extension}

In this section, we
briefly discuss the spacetime meaning of the leg factor
in the high-energy limit and try to interpret
our result qualitatively. Consider the
in($+$) and out($-$) string-wave packets which are
localized with respect to
their center of mass,
\eqn\pack{
\psi_{\pm}(X\pm \varphi)
 = \int {d\omega \over 2\pi\omega} f_{\pm}(\omega)
e^{-i\sqrt{2}\omega (X\pm\varphi)}.}
If the modulus $\vert f_{\pm}(\omega) \vert$ of the function
$f_{\pm}(\omega)$ is sharply peaked at $\omega=\omega_0$,
the trajectory of the wave packet is given,
using the phase $\phi_{\pm}(\omega)$
of the function $f_{\pm}(\omega)$, as
\eqn\traj{
X= \mp \varphi + \delta_{\pm}(\omega_0),
}
where
\eqn\delay{
\delta_{\pm}(\omega_0) \equiv
 {1\over \sqrt{2}}{\partial \phi_{\pm}(\omega) \over
\partial \omega}\bigr\vert_{\omega=\omega_0}.
}
The general transition matrix element for the wave packet
states is
\eqn\smatrix{\eqalign{
S_{nm} =
&\bigl(\prod_{i=1}^n \int {d\omega_i \over 2\pi\omega_i}\bigr)
\bigl(\prod_{j=n+1}^{n+m} \int {d\omega_{j} \over 2\pi\omega_j}\bigr)
\delta(\sum_{i=1}^n \omega_i-\sum_{j=1}^m\omega_{n+j}) \cr
&\times (\prod_{i=1}^n \overline{f_{i-}(\omega_i)})
S(\omega_1,\omega_2, \ldots, \omega_n
; -\omega_{n+1}, \ldots, -\omega_{n+m})\cr
&\times (\prod_{j=1}^m f_{j+}(\omega_{n+j})).\cr}}

The trajectories of the scattered wave packets are determined
by the condition that the phase of the integrand in eq.\smatrix\
is stationary with respect to the variation of independent
energies $\omega_i$'s. Since the matrix model
$S$-matrix element $\bar S(\omega_i)$ has no phase factor,
this means that the trajectories are determined by the
leg factor. In the high energy limit studied here
\foot{
In the early-time limit discussed in \np,
the trajectories are determined by the behavior of
the  phase at the first pole on the imaginary
axis. Note that the early-time limit
is necessarily a low-energy approximation.
In the present paper, we are interested in the high-energy
limit and the phase is governed by the collective effect of all poles.
},
they are thus obtained by solving
\eqn\dettraj{
{\partial \over \partial \tilde \omega_a}\{\sum_{i=1}^n
(-\phi_{i-}(\omega_i)+\gamma(\omega_i))+
\sum_{j=1}^{n+j}(\phi_{j+}(\omega_{j+n})
+\gamma(\omega_{n+j}))\}=0,
}
where $\{\tilde \omega_a : a=1, 2, \ldots, n+m-1\}$ denotes the
independent set of energies, $\phi_{i-}, \,
\phi_{j+}$ the phases of the functions
$f_{i-}, f_{j+}$, respectively, and
\eqn\pshift{
\gamma(\omega) = 8\omega(\ln 2\omega -1)-2\omega \ln \mu.
}
Here and what follows, we drop the indices $0$ for the
peaked values of energies.

For example, in the simplest ($1\rightarrow 1$) scattering
$n=m=1$ with $\omega_1=\omega_2=\omega$, we have actual time delay
\eqn\twopdelay{
\delta_{1-}-\delta_{1+} = {d\gamma(\omega)\over d\omega}
= 8\ln 2\omega -2\ln \mu.
}
As we have emphasized in previous sections,
the first term is twice that of ordinary
point particle in Liouville potential.

For ($1\rightarrow 2$) scattering with
$\omega_3=\omega=\omega_1+\omega_2$, the conditions
\dettraj\ are
\eqn\threedecay{
{\omega_1\over \omega}\delta_{1-}
+{\omega_2 \over \omega}\delta_{2-}
-\delta_{3+}
={d\gamma (\omega)\over d\omega}
+\sum_{i=1}^2 {\omega_i\over \omega}
{d\gamma (\omega_i)\over d\omega_i},
}
\eqn\threedecayrel{
\delta_{1-}-\delta_{2-}
={d\gamma (\omega_1)\over d\omega_1}
-{d\gamma (\omega_2)\over d\omega_2}.
}
Similarly, for ($2\rightarrow 1$) scattering with
$\omega_1=\omega=\omega_2+\omega_3$, we have
\eqn\threesca{
\delta_{1-}-({\omega_1\over \omega}\delta_{1+}+
{\omega_2 \over \omega}\delta_{2+})
={d\gamma (\omega)\over d\omega}
+\sum_{i=2}^3 {\omega_i\over \omega}
{d\gamma (\omega_i)\over d\omega_i},
}
\eqn\threeprel{
\delta_{1+}-\delta_{2+}
=-{d\gamma (\omega_2)\over d\omega_2}
+{d\gamma (\omega_3)\over d\omega_3}.}
The trajectory of each wave packet is obtained by
plugging $\delta_{\pm}(\omega_i)$ into \traj.
 From \threedecay\  and \threesca, we see,
in the extreme high-energy limit $\omega\rightarrow
\infty$ where we have ${d\gamma(\omega) \over d\omega}\sim
8\ln \omega$, that
the average time delay between initial and final
wave packets is always given by the result of the two-point
scattering. In this qualitative sense, doubling of
phase shift is universal for arbitrary scattering of 2D strings
in tachyon background.

The computations of previous sections clearly show
that the additional contribution
to the time delay arises from the extension
of strings. This is natural
because the duration of interaction of a string with a given potential
would in general be longer, owing to string extension,
than for a point particle of the same velocity.
Let us try to qualitatively estimate the time delay
arising from string extension.
In the case of ordinary critical strings,
it has been argued by several authors \ven\gm\yo\  that the string
extension $\triangle \ell$
increases in proportion to its energy $\omega$
in the high energy limit. We can interpret
this as follows. As the energy of a string increases,
more and more energy can be exchanged in the process of interaction
between the center of mass and string excitation.
A large fluctuation in excitation energy implies
a large string extension, which is estimated to be
proportional to the energy \yo.
In our two-dimensional case, there are no transverse excitations
in the usual sense. However, there still appears an infinite sequence
of discrete excited states at discrete imaginary momenta.
In the high-energy limit, we naturally suppose that all these discrete states
can collectively participate in the fluctuation of energies.
If the fluctuation of the excitation
energy  and hence of the center of mass energy
is of order $\triangle E \sim \omega$ during interaction,
it is expected, by converting it to uncertainty
with respect to the $\varphi$ coordinate in the Liouville potential
$e^{-\sqrt{2}\varphi}$,
to give an additional contribution to
the time delay of order
$\ln \triangle E \sim \ln \omega \, (\sim \ln \triangle \ell)$.

To summarize,
We have given a new direct derivation of the
phase shift for 2-point wall scattering amplitude
in the high-energy limit
and suggested a general scheme for extending our
calculation to higher-point amplitudes.
The result indicates that the high-energy behavior
of the leg factor reflects the effect of string
extension in a given background. We have suggested that the
additional contribution to the time delay is due to
the large energy fluctuation occuring between the
center of mass and discrete excited states in the
high-energy limit.

Next immediate problem would be to extend the
result to the case of black hole background.
If our interpretation is correct,
the high-energy limit of the leg factor
in this case should also exhibit properties which are
qualitatively the same as in the tachyon background.
We are planning to report about this in a separate
publication.

\medskip
\noindent{\bf Acknowledgments}

The present collaboration was made possible by the US-Japan collaborative
program in Science
conducted by NSF and JSPS. We would like to thank NSF and JSPS
for their financial support. The research of A. J. and M. L. was supported
by DOE Contract No. DE-FG02-91ER40688-Task A. A.J. would like to thank the
Institute of Physics of the University of Tokyo at Komaba campus
 for its kind hospitality.
T. Y. would like to thank the Physics Department of Brown University
for its kind hospitality. The research of T. Y.
was partially supported by the Grant-in-Aid for Scientific Research
(No. 06640378) and Grant-in-Aid for Priority Area (No. 06221211) from
the Ministry of Education, Science, and Culture.

\vfill
\eject

\listrefs
\end